# Time reversal and negative energies in general relativity

J. M. Ripalda[*]

The significance of past-pointing four-vectors and negative energies in general relativity is discussed. The sign of the energy is not absolute, but relative to the four-velocity of the observer, and every particle/observer always measures its own mass as positive. It is shown that the description of the interaction of past-pointing and future-pointing matter in general relativity requires two metric tensors for self-consistency. This aspect of general relativity might account for the observations that led to the proposal of "dark energy" and non-baryonic "dark matter".

The current difficulties with "dark energy", "dark matter", large scale flatness, the near zero value of the cosmological constant, and the non-renormalizability of quantum gravity, might be related to the current views on time reversal symmetry and the significance of negative energies.

It has many times been argued that negative energies are to be ruled out as they would lead to vacuum decay into zero total energy-momentum particle pairs. Various authors have calculated the particle creation rate due to vacuum decay under various assumptions.[1,2,3] There is no reason for this process to occur only in one direction (creation), the time reversed process (annihilation) having the same transition probability amplitude due to the symmetries of the scattering matrix. Furthermore, for every contribution to the fields from the vacuum decay processes, there would be an exactly opposite contribution interfering destructively. The resulting fields would have zero total energy-momentum and zero total charge. Vacuum decay should not be a problem if it does not have an effect on observable expectation values. Discarding negative energy states brings up a problem no less difficult than vacuum decay, leaving nothing to compensate the positive contributions to the zero-point energy of vacuum.

The idea of vacuum decay and the fact that, experimentally, particle-antiparticle pairs have a total mass of $2m$, not zero, led to the general acceptance of the quantum mechanical time reversal operator being defined as the product of the unitary time reversal operator and the complex conjugation operator, so energies would remain positive under time reversal. In general relativity, energy, mass, and the stress-energy tensor are unchanged under a global reversal of the coordinate system (PT), but reversal of the proper time of a single world-line (reversing a future-pointing four-vector into a past-pointing four-vector), inevitably leads to negative energies (see Costella *et al.* for an attempt to avoid negative energies in general relativity[4]). It is customary to use a variety of energy conditions (strong, weak, null, dominant, averaged null, *etc*) to define which of the solutions of general relativity are to be rejected as unphysical,[5,6] and past-pointing four-vectors are commonly regarded as not having much physical significance other than as a mathematical tool that can be used to describe charge conjugated matter. But the equations of classical physics show no preferred direction in time. The possibility of particles propagating backwards in time is inherent to all relativistic theories, quantum or not. Wheeler and Feynman found that a consistent description of radiative reaction in classical physics requires the usage of past-pointing four-vectors ("advanced action") on an equal footing to future-pointing four-vectors ("retarded action").[7]

In the first half of this paper the effect of discrete symmetries will be deduced from the equations of general relativity, and their relevance to astronomical and experimental observations will be discussed in the second part of the paper.

The "length" of a particle's worldline is parameterized by its proper time, defined as:

$$d\tau = \pm\sqrt{dx^\mu dx_\mu}$$

Solutions to the square root above with a positive or a negative sign are equally valid. From the Lorentz transformation in flat space-time:

$$d\tau = \frac{dt'}{\gamma} \quad (1)$$

where

$$\gamma = \frac{1}{\pm\sqrt{1 - v^2/c^2}} \quad (2)$$

$$u^\mu = \gamma(1, \mathbf{u}') = \frac{\partial x^\mu}{\partial \tau} \quad (3)$$

j.ripalda@gmail.com

The square root in the denominator of $\gamma$ is again a reminder that $\tau$ and $\gamma$ can either take a positive or a negative sign.

Both PT and $\tau \rightarrow -\tau$ reverse four-velocities and four-momenta, while only PT reverses event positions and accelerations, as these are even in $\tau$. The PT reversal of the coordinate system affects all particles and observers, while proper time is a property intrinsic to each worldline, and thus $\tau \rightarrow -\tau$ only affects the worldline in question.

The equivalence principle states that the outcome of any local experiment in a free falling laboratory is independent of the velocity of the laboratory and its location in space-time. In general relativity, velocities can be past-pointing or future-pointing four-vectors, as the theory is perfectly symmetric in time. Therefore the equivalence principle also implies that the laws of physics are the same for future-pointing ($\tau^+$) and past-pointing ($\tau^-$) observers, so that the terms "past" and "future" do not have an absolute meaning, they only have a meaning relative to the reference frame of the observer. A past-pointing observer in a past-pointing world observes the same phenomena as we do. Both past-pointing and future-pointing four-vectors are equally valid, and both can coexist in the same reference frame (e.g.: both are needed in the classical theory of radiative reaction[7]), so this is not merely a matter of convention in the definition of the sign of time and energy. Quoting Wheeler and Feynman: "Generalizing, we conclude advanced and retarded interactions give a description of nature logically as acceptable and physically as completely deterministic as the Newtonian scheme of mechanics. In both forms of dynamics the distinction between cause and effect is pointless."[7]

Consider the energy density $E$ measured by an observer at some point of a massive object:

$$E = p^\mu u_\mu = T^{\mu\nu} u_\mu u_\nu \qquad (4)$$

where $p$ is the four-momentum density of the object, and $u$ is the four-velocity of the observer. From the equation above it is inevitable that the scalar energy and stress-energy tensor $T$ change sign when the object has its four-momentum reversed or the observer has its four-velocity reversed. If the proper times of both the object and the observer are reversed, then the energy does not change sign. There is no reason to expect an spontaneous decay from positive to negative energy states because the sign of the energy is not absolute, but relative to the four-velocity of the observer, and every particle/observer always measures its own mass as positive. The change in the sign of the stress-energy tensor obtained in equation (4) from a change in sign of either $u$ or $p$ (but not both), implies that the geometry of space-time measured by a past-pointing observer is not the same as the geometry of space-time measured by a future-pointing observer. The stress-energy $T$, being a rank 2 tensor, is even under a reversal of the space-time coordinate system (PT), but as shown above, odd under $\tau \rightarrow -\tau$. It might be argued that this is a violation of the principle of uniqueness of free fall, but the geodesic is only uniquely defined for a set of initial conditions that specify an event and a four-velocity, and reversing the proper time is a reversal of the four-velocity, and thus a change in the state of motion.

Another hint at the inevitability of a change of sign in $T$ for past-pointing sources/observers results from the following equation for the observed mass density as a function of the four-velocity of the observer:

$$m = \frac{1}{c^2} T^{\mu\nu} u_\mu u_\nu = \gamma\, m_0 \qquad (5)$$

The choice of sign for the proper time $\tau$ of the observer is linked by equation 1 to the sign of $\gamma$, by equation 3 to the sign of the four-velocity of the observer, and by equation 5 to the sign of the observed mass-energy scalar and stress-energy tensor $T$. The contribution of past-pointing $\tau^-$ matter to the stress-energy tensor $T$ has the opposite sign than the contribution from future-pointing $\tau^+$ matter. A change of sign in $T$ has implications on the geometry of space-time as:

$$R_{\mu\nu} - \frac{1}{2} R g_{\mu\nu} = \frac{8\pi k}{c^4} T_{\mu\nu}$$

where $g$ is the space-time metric, $T$ is the stress-energy tensor, $R_{\mu\nu}$ is the Ricci contraction of the Riemann tensor of curvature that describes the geometry of space-time, and $R$ is the scalar curvature. From the equation above follows that a change of sign in $T$ changes the Riemann tensor of curvature, the metric, and the geodesic curves in a non-trivial way. Hossenfelder has proposed a double metric for general relativity, each metric corresponding to source fields of positive and negative mass.[8] From the discussion above it follows that when either the source or the observer is represented by a past-pointing four vector, the stress-energy tensor to be used has the opposite sign than the conventional one, and consequently a second metric is required for consistency. Hossenfelder makes a distinction between inertial and gravitational mass, distinction which is not made here. Related ideas have been proposed by Nicker and Henry-Couannier, among others.[9,10]

Summarizing the effect of discrete symmetries on physical quantities:

|  | $\tau \to -\tau$ | PT |
|---|---|---|
| $\gamma$ | - | + |
| $m$ | - | + |
| $x^\mu$ | + | - |
| $u^\mu$ | - | - |
| $p^\mu$ | - | - |
| $\partial u^\mu/\partial \tau$ | + | - |
| $T^{\mu\nu}$ | - | + |

Given the perfect symmetry in time of the equations of general relativity, it seems natural to assume that there is the same amount of matter moving forwards in time ($\tau^+$) as of matter moving backwards in time ($\tau^-$), thus $\tau^-$ matter would compensate the space-time curvature induced by $\tau^+$ matter, in agreement with recent observations suggesting that the universe is nearly flat on a large scale.[11] Due to the gravitational repulsion between $\tau^+$ and $\tau^-$ matter, gravity must have expelled most $\tau^-$ matter to cosmological distances away from us, thus we cannot expect earthbound experiments and observations of nearby galaxies to yield much information about the physics of $\tau^-$ matter. It is tempting to think of the gravitational repulsion between $\tau^+$ and $\tau^-$ matter as a possible explanation for the matter-antimatter asymmetry in our proximity, but future-pointing positrons would still be gravitationally attracted to future-pointing electrons. According to the Feynman-Stueckelberg interpretation of antimatter, charge conjugated matter ($q^-\tau^+$) and past-pointing matter ($q^+\tau^-$) are indistinguishable in flat space-time. But such equivalence has not been proved in any context including the effects of gravity. Classical physics is separately invariant under each of the C, P, and T symmetries, and consequently also under the combined CPT symmetry, but that does not imply $q^-\tau^+$ and $q^+\tau^-$ being indistinguishable (just like the P symmetry of the laws of classical physics does not imply the indistinguishability of left handed and right handed matter). Thus the term "antimatter" might not be sufficiently specific in the context of general relativity. Four types of matter/antimatter are conceivable, labelled as $q^+\tau^+$, $q^+\tau^-$, $q^-\tau^+$, $q^-\tau^-$, being combinations of matter with reversed signs for their charges and proper times. A $q^+\tau^+/q^+\tau^-$ particle pair has zero total mass, whereas a $q^+\tau^+/q^-\tau^+$ particle pair has a total mass of $2m$, which is the type of particle pair observed in earthbound experiments as, by definition, matter on earth is of the $\tau^+$ type. Despite the success of the Feynman-Stueckelberg interpretation of antimatter in all theories that do not include the effects of gravity, we are forced to conclude that it does not work in the context of general relativity. Gravity breaks the Feynman-Stueckelberg equivalence of $q^+\tau^-$ and $q^-\tau^+$ matter.

The various theoretical difficulties that led to the early rejection of the idea of repulsive gravity have been critically reviewed by Nieto et al.[12] In 1961, before the experimental discovery of CP violation, Myron Good argued against the gravitational repulsion between matter and antimatter that this would have as a consequence a process known in particle physics as neutral Kaon decay, and such process had not been observed at the time.[13] With the exception of Chardin,[14,15] few authors have paid attention to the predictions of Good after neutral Kaon decay was experimentally observed.

The idea of a universe with matter and antimatter domains was studied by Brown and Stecker.[16] These authors suggested that grand unified field theories with spontaneous symmetry breaking in the early big bang could lead more naturally to a baryon-symmetric cosmology with a domain structure than to a baryon-asymmetric cosmology. Alfven also studied a similar cosmological model.[17]

The brightness vs. red shift statistics of supernovae suggests that the velocity of expansion of the universe is not being gradually slowed down by gravity, but much to the contrary is being accelerated by some unknown repulsive force.[18,19] This is most often interpreted in terms of Einstein's cosmological constant or "dark energy". Perhaps these observations can also be interpreted in terms of the gravitational repulsion between $\tau^-$ and $\tau^+$ matter. A universe with equal amounts of $\tau^-$ and $\tau^+$ matter clusters might imply the possibility of large scale annihilation resulting in gamma ray bursts, but repulsive gravitational interaction between $\tau^-$ and $\tau^+$ matter would reduce the likelihood of such events. The precise location of the sources of Gamma Ray Bursts and rough estimations of the total emitted energy has become possible only recently (1998). The energy liberated in some of these events is of the order of the rest mass of two stars with the size of the sun.[20,21]

For illustrative purposes we discuss bellow a very simple model for the net force on a universe with large scale $\tau^-$ and $\tau^+$ matter symmetry. If there is the same amount of $\tau^-$ and $\tau^+$ matter the forces do not necessarily cancel out. In an ionic solid there is the same number of positive and negative charges, but the overall electrostatic force on the crystal is attractive (compensated by the Fermi exclusion principle). If we now visualize each positive ion as a cluster of $\tau^+$ matter galaxies and each negative ion as a $\tau^-$ matter cluster, and replace the electrostatic interaction by the gravitational potential due to a point mass in the Newtonian limit, we obtain from the Madelung model of an ionic solid:

$$U_g = \frac{1}{2} N\alpha \frac{m^2}{R}$$

where $U_g$ is the total gravitational energy, $N$ is the total number of clusters, $m$ would represent the mass of the clusters, $R$ the separation between nearest neighbors, and $\alpha$ is the Madelung constant. For simplicity, and to keep the analogy with a crystal, $m$ and $R$ are assumed to be the same for all clusters. The Madelung constant takes values between 1.8 and 1.6 for most crystal structures. The overall force on the universe ($dU_g / dR$) is repulsive. Such a model of the universe could not be static, the effective values of $R$, $m$, and $\alpha$ being a function of time. Clusters of the same sign would have a tendency to coalesce, so $N$ would decrease with time, while $R$ and $m$ would increase. Gravitational forces would only be perfectly compensated in a homogeneous universe. In such a universe, gravity would amplify initial inhomogeneities due to quantum fluctuations and eventually this would lead to the formation of large clusters of $\tau^+$ and $\tau^-$ matter. Therefore an initially at rest and homogeneous universe would spontaneously evolve into an inhomogeneous universe in accelerated expansion.

One of the more intriguing aspects of "dark matter" is that its apparent distribution is different from the distribution of matter. How could this be if "dark matter" interacts gravitationally and follows the same geodesics as all matter and energy? A possible explanation is suggested by an analogy with electrostatics: the potential created by a cluster of positive charges is exactly equivalent to the potential created by an empty region in an otherwise homogenous distribution of negative charges. An example is the case of holes in a semiconductor: these are empty electron states that are modelled as positively charged particles. Any large cluster of $\tau^+$ matter can be expected to be surrounded by a region void of $\tau^-$ matter due to the gravitational repulsion between $\tau^+$ and $\tau^-$ matter. A void in a homogenous distribution of $\tau^-$ matter creates a gravitational potential that is equivalent to the potential created by halo of $\tau^+$ matter. Therefore we can expect every galaxy to be surrounded by a gravitational halo of apparent "dark matter" that in fact is just a void in the distribution of $\tau^-$ matter.

Summarizing: although the Feynman-Stueckelberg interpretation of antimatter is perfectly consistent with quantum and classical electrodynamics, it is here argued that such interpretation is not consistent with general relativity. Past-pointing four-vectors cannot be dealt with consistently in general relativity unless we accept that the stress-energy tensor created/observed by matter going backwards in time inevitably has a sign opposite to the stress-energy tensor created/observed by matter going forwards in time. Therefore the geometry of space-time has two metric tensors, two branches. Time reversed matter moves on a different branch than future pointing matter. Both branches are related by sharing the same stress-energy tensor up to a change in sign, and so they interact through repulsive gravity. On a large scale, equal amounts of future-pointing and past-pointing matter should be expected. Due to their mutual repulsion, there should be voids in the distribution of past-pointing matter around future-pointing clusters. Such voids create an effective "dark" gravitational halo around matter clusters, just like a hole creates the effect of a positive charge in a semiconductor. The concepts of "dark energy" and non-baryonic "dark matter" are unnecessary. The fact that we experience time as always going forwards is due to the separation of past-pointing matter and future-pointing matter by gravity (a spontaneous local symmetry breaking). On a large scale, there is no "arrow of time".